\documentclass[]{aa}

\usepackage{txfonts}
\usepackage{graphicx}
\usepackage{natbib}
\bibpunct{(}{)}{;}{a}{}{,}

\begin{document}

\title{The ALHAMBRA survey\thanks{Based on observations collected at the German-Spanish Astronomical Center, Calar Alto, jointly operated by the Max-Planck-Institut f\"ur Astronomie (MPIA) at Heidelberg and the Instituto de Astrof\'{\i}sica de Andaluc\'{\i}a (CSIC)}: tight dependence of the optical mass-to-light ratio on galaxy colour up to $z = 1.5$}

\author{C.~L\'opez-Sanjuan\inst{1}
\and L.~A.~D\'{\i}az-Garc\'{\i}a\inst{1}
\and A.~J.~Cenarro\inst{1}
\and A.~Fern\'andez-Soto	\inst{2,3}
\and K.~Viironen\inst{1}
\and A.~Molino			\inst{4,5}
\and N.~Ben\'{\i}tez		\inst{5}
\and D.~Crist\'obal-Hornillos	\inst{1}
\and M.~Moles			\inst{1}
\and J.~Varela                  \inst{1}
\and P.~Arnalte-Mur		\inst{6,7}
\and B.~Ascaso			\inst{8}
\and F.~J.~Castander		\inst{9,10}
\and M.~Cervi\~no		\inst{5,11}
\and R.~M.~Gonz\'alez Delgado	\inst{5}
\and C.~Husillos		\inst{5}
\and I.~M\'arquez		\inst{5}
\and J.~Masegosa		\inst{5}
\and A.~Del Olmo		\inst{5}
\and M.~Povi\'c			\inst{12,5}
\and J.~Perea			\inst{5}
}

\institute{Centro de Estudios de F\'{\i}sica del Cosmos de Arag\'on, Unidad Asociada al CSIC, Plaza San Juan 1, 44001 Teruel, Spain\\
\email{clsj@cefca.es}
\and Instituto de F\'{\i}sica de Cantabria (CSIC-UC), 39005, Santander, Spain 
\and Unidad Asociada Observatorio Astron\'omico (IFCA-UV), 46980, Paterna, Spain 
\and Universidade de S\~ao Paulo, Instituto de Astronomia, Geof\'{\i}sica e Ci\^encias Atmosf\'ericas, Rua do Mat\~ao 1226, 05508-090, S\~ao
Paulo, Brazil 
\and Instituto de Astrof\'{\i}sica de Andaluc\'{\i}a (IAA-CSIC), Glorieta de la astronom\'{\i}a s/n, 18008 Granada, Spain 
\and Observatori Astron\`omic, Universitat de Val\`encia, C/ Catedr\'atico Jos\'e Beltr\'an 2, 46980 Paterna, Spain 
\and Departament d'Astronomia i Astrof\'{\i}sica, Universitat de Val\`encia, 46100 Burjassot, Spain 
\and APC, AstroParticule et Cosmologie, Universit\'e Paris Diderot, CNRS/IN2P3, CEA/lrfu, Observatoire de Paris, Sorbonne Paris Cit\'e, 10, rue Alice Domon et L\'eonie Duquet, 75205 Paris Cedex 13, France 
\and Institut of Space Sciences (ICE, CSIC), Campus UAB, Carrer Can Magrans, s/n, 08193 Barcelona, Spain 
\and Institut d'Estudis Espacials de Catalunya (IEEC), 08193 Barcelona, Spain 
\and Instituto de Astrof\'{\i}sica de Canarias, V\'{\i}a L\'actea s/n, La Laguna, 38200 Tenerife, Spain 
\and Ethiopian Space Science and Technology Institute (ESSTI), Entoto Observatory and Research Center (EORC), Astronomy and Astrophysics Research Division, PO Box 33679, Addis Ababa, Ethiopia 
}

\date{Submitted --, 2018}

\abstract
{}
{Our goal is to characterise the dependence of the optical mass-to-light ratio on galaxy colour up to $z = 1.5$, expanding the redshift range explored in previous work.}
{From the ALHAMBRA redshifts, stellar masses, and rest-frame luminosities provided by the \texttt{MUFFIT} code, we derive the mass-to-light ratio vs. colour relation (MLCR) both for quiescent and star-forming galaxies. The intrinsic relation and its physical dispersion are derived with a Bayesian inference model.}
{The rest-frame $i-$band mass-to-light ratio of quiescent and star-forming galaxies presents a tight correlation with the rest-frame $(g-i)$ colour up to $z = 1.5$. Such MLCR is linear for quiescent galaxies and quadratic for star-forming galaxies. The intrinsic dispersion in these relations is 0.02 dex for quiescent galaxies and 0.06 dex for star-forming ones. The derived MLCRs do not present a significant redshift evolution and are compatible with previous local results in the literature. Finally, these tight relations also hold for $g-$ and $r-$band luminosities.}
{The derived MLCRs in ALHAMBRA can be used to predict the mass-to-light ratio from a rest-frame optical colour up to $z = 1.5$. These tight correlations do not change with redshift, suggesting that galaxies have evolved along the derived relations during the last 9 Gyr.}


\titlerunning{The ALHAMBRA survey. Optical mass-to-light ratio vs. colour up to $z = 1.5$}

\authorrunning{L\'opez-Sanjuan et al.}

\maketitle

\section{Introduction}\label{intro}
Stellar mass is a fundamental parameter in galaxy evolution studies, presenting correlations with several galaxy properties such as star formation rate \citep[e.g.][]{noeske07,chang15}, gas-phase metallicity \citep[e.g.][]{tremonti04,mannucci09,lara10}, stellar content \citep[e.g.][]{gallazzi05,gallazzi14,diazgarcia18sp}, galaxy size \citep[e.g.][]{shen03,trujillo07,vanderwel14}, morphology \citep[e.g.][]{moffett16,huertas16}, or nuclear activity \citep[e.g.][]{kauffmann03,bongiorno16}.

The measurement of stellar mass in modern photometric and spectroscopic surveys is mainly performed by comparing either an empirical or a theoretical library of templates with the observational spectral energy distribution (SED) of galaxies. The mass-to-light ratio associated to the templates, combined with the flux normalization, provides the stellar mass of a given source \citep[see][for a recent review on galaxy mass estimation]{courteau14}. Thus, understanding and characterising the mass-to-light ratio of different galaxy populations is important to derive reliable stellar masses as well as to minimise systematic differences between data sets and template libraries.

The mass-to-light versus colour relations (MLCRs) have been studied theoretically and observationally \citep{tinsley81, jablonka92, bell01, bell03, portinari04, gallazzi09, zibetti09, taylor11, into13, mcgaugh14, zaritsky14, vandesande15, roediger15, herrmann16} in the optical, the ultraviolet (UV), and the near-infrared (NIR). These studies find well defined linear MLCRs with low scatter (< 0.2 dex) and focus in the low redshift Universe ($z \lesssim 0.5$).

We highlight the work of \citet[][T11 hereafter]{taylor11}. It is based in the SED-fitting to the $ugriz$ broad bands of the Sloan Digital Sky Survey (SDSS DR7, \citealt{sdssdr7}) available for the GAMA \citep[Galaxy And Mass Assembly, ][]{gama} survey area. They find a remarkable tight relation (0.1 dex dispersion) between the mass-to-light ratio in the $i$ band, noted $M_{\star}/L_{i}$, and the rest-frame colour $(g-i)$ at $z < 0.65$, with a median redshift of $\langle z \rangle = 0.2$ for the analysed global population. T11 argue that this small dispersion is driven by (i) the degeneracies of the galaxy templates in such a plane, that are roughly perpendicular to the MLCR, implying from the theoretical point of view $\sim$0.2 dex errors in the mass-to-light ratio even with large errors in the derived stellar population parameters. And (ii) the galaxy formation and evolution processes, that are encoded in the observed galaxy colours and only allow a limited set of solutions, making the observed relation even tighter than the theoretical expectations.

In the present work, we expand the results from T11 with the multi-filter ALHAMBRA\footnote{\tt www.alhambrasurvey.com} (Advanced, Large, Homogeneous Area, Medium-Band Redshift Astronomical) survey \citep{alhambra}. ALHAMBRA provides stellar masses thanks to the application of the Multi Filter FITing (\texttt{MUFFIT}, \citealt{diazgarcia15}) code to 20 optical medium-band and 3 NIR photometric points. In addition, ALHAMBRA covers a wide redshift range, reaching $z = 1.5$ with a median redshift of $\langle z \rangle = 0.65$, and reliably classifies quiescent and star-forming galaxies thanks to dust de-reddened colours.

We also refine the statistical estimation of the MLCRs. Instead of performing an error-weighted fit to the data, we applied a Bayesian inference model that accounts for observational uncertainties and includes intrinsic dispersions in the relations \citep[see][for other applications of such kind of modelling]{taylor15,monterodorta16lf}.

The paper is organised as follows. In Sect.~\ref{data}, we present the ALHAMBRA photometric redshifts, stellar masses, and luminosities. The derived $i-$band MLCRs for quiescent and star-forming galaxies and their modelling are described in Sect.~\ref{mlratio}. Our results are presented and discussed in Sect.~\ref{results}. Summary and conclusions are in Sect.~\ref{conclusions}. Throughout this paper we use a standard cosmology with $\Omega_{\rm m} = 0.3$, $\Omega_{\Lambda} = 0.7$, $\Omega_{\rm k} = 0$, $H_{0}= 100h$ km s$^{-1}$ Mpc$^{-1}$, and $h = 0.7$. Magnitudes are given in the AB system \citep{oke83}. The stellar masses, $M_{\star}$, are expressed in solar masses ($M_{\odot}$) and the luminosities, $L$, in units equivalent to an AB magnitude of 0. The derived mass-to-light ratios can be transformed into solar luminosities $L_{\odot}$ by subtracting 2.05, 1.90, and 1.81 to the presented MLCRs for the $g$, $r$, and $i$ bands, respectively. With the definitions above, stellar masses can be estimated from the reported mass-to-light ratios as
\begin{equation}
\log_{10} M_{\star} = \log_{10}\,(M_{\star}/L) - 0.4M,
\end{equation}
where $M$ is the absolute AB magnitude of the galaxy.


\section{ALHAMBRA survey}\label{data}
The ALHAMBRA survey provides a photometric data set over 20 contiguous, equal-width ($\sim$300\AA), non-overlapping, medium-band optical filters (3500\AA - 9700\AA) plus 3 standard broad-band NIR filters ($J$, $H$, and $K_{\rm s}$) over 8 different regions of the northern sky \citep{alhambra}. The final survey parameters and scientific goals, as well as the technical properties of the filter set, were described by \citet{alhambra}. The survey collected its data for the 20+3 optical-NIR filters in the 3.5m telescope at the Calar Alto observatory, using the wide-field camera LAICA (Large Area Imager for Calar Alto) in the optical and the OMEGA–2000 camera in the NIR. The full characterisation, description, and performance of the ALHAMBRA optical photometric system was presented in \citet{aparicio10}. A summary of the optical reduction can be found in \citet{molino13}, while that of the NIR reduction is in \citet{cristobal09}.

\subsection{Bayesian photometric redshifts in ALHAMBRA}
The Bayesian photometric redshifts ($z_{\rm b}$) of ALHAMBRA were estimated with \texttt{BPZ2}, a new version of the Bayesian photometric redshift (\texttt{BPZ}, \citealt{benitez00}) code. The \texttt{BPZ2} code is a SED-fitting method based in a Bayesian inference, where a maximum likelihood is weighted by a prior probability. The template library comprises 11 SEDs, with four ellipticals, one lenticular, two spirals, and four starbursts. The ALHAMBRA photometry used to compute the photometric redshifts is PSF-matched aperture-corrected and based on isophotal magnitudes \citep{molino13}. In addition, a recalibration of the zero point of the images was performed to enhance the accuracy of the photometric redshifts. Sources were detected in a synthetic $F814W$ filter image defined to resemble the HST/$F814W$ filter. The total area covered by the current release of the ALHAMBRA survey after masking low signal-to-noise areas and bright stars is 2.38 deg$^{2}$ \citep{arnaltemur14}. The full description of the photometric redshift estimation is detailed in \citet{molino13}.

The photometric redshift accuracy, as estimated by comparison with spectroscopic redshifts ($z_{\rm s}$), is $\sigma_{\rm NMAD} = 0.012$ at $F814W \leq 23$. The variable $\sigma_{\rm NMAD}$ is the normalized median absolute deviation of the photometric vs. spectroscopic redshift distribution \citep[e.g.][]{ilbert06, molino13}. The fraction of catastrophic outliers with $|z_{\rm b} - z_{\rm s}|/(1 + z_{\rm s}) > 0.2$ is 2.1\%. We refer to \citet{molino13} for a more detailed discussion.

\begin{figure}[t]
\centering
\resizebox{\hsize}{!}{\includegraphics{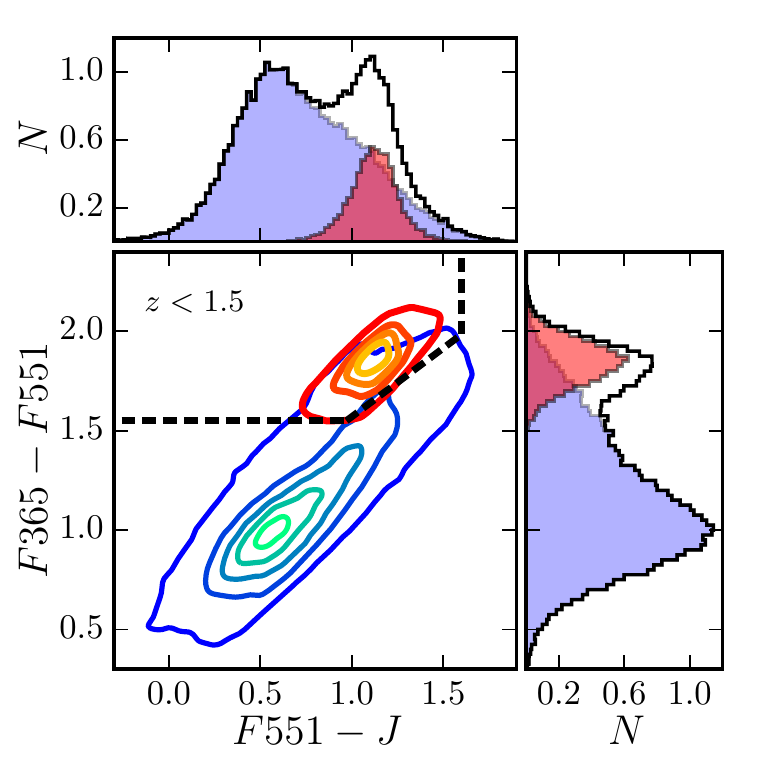}}
\caption{Rest-frame colour-colour plane $F365 - F551$ vs. $F551 - J$ for the 76642 ALHAMBRA galaxies with $F814W \leq 23$ at $z < 1.5$. This plane is equivalent to the commonly used $UVJ$ diagram, and the quiescent selection box from the literature \citep{williams09} is delimited by dashed lines. The coloured lines show level contours in the density of galaxies, starting at 0.1 galaxies dex$^{-2}$ and increasing in 0.6 galaxies dex$^{-2}$ steps. Red contours show the quiescent population, and blue contours show the star-forming one, as defined with dust-corrected colours by DG17. The side panels show the normalized distribution in $F365 - F551$ ({\it right panel}) and $F551 - J$ ({\it top panel}). In both panels the total distribution is presented in black, the quiescent population in filled red, and the star-forming population in filled blue.}
\label{UVJ}
\end{figure}

\subsection{\texttt{MUFFIT}: stellar masses and rest-frame colours}\label{muffit}
The \texttt{BPZ2} template library presented above is empirical, and the different templates have not assigned mass-to-light ratios {\it a priori}. Hence, an alternative methodology is needed to compute the stellar mass of the ALHAMBRA sources.

The \texttt{MUFFIT} code is specifically performed and optimized to deal with multi-photometric data, such as the ALHAMBRA dataset, through the SED-fitting (based in a $\chi^2$-test weighted by errors) to mixtures of two single stellar populations (a dominant ``old'' component plus a posterior star formation episode, which can be related with a burst or a younger/extended tail in the star formation history). \texttt{MUFFIT} includes an iterative process for removing those bands that may be affected by strong emission lines, being able to carry out a detailed analysis of the galaxy SED even when strong nebular or active galactic nuclei (AGN) emission lines are present, which may be specially troublesome for intermediate and narrow band surveys. ALHAMBRA sources with $F814W \leq 23$ are analysed with \texttt{MUFFIT} by \citet[][hereafter DG17]{diazgarcia18uvj}, retriving ages, metallicities, stellar masses, rest-frame luminosities, and extinctions. \texttt{MUFFIT} also provides photometric redshifts, using the \texttt{BPZ2} solutions presented in the previous section as a prior to minimise degeneracies and improving the photometric redshift accuracy by $\sim20$\%. The retrieved parameters are in good agreement with both spectroscopic diagnostics from SDSS data and photometric studies in the COSMOS survey with shared galaxy samples (\citealp{diazgarcia15}, DG17). 

To study the MLCR of ALHAMBRA galaxies and its redshift evolution, we used the redshifts, stellar masses, and rest-frame luminosities in the $gri$ broad-bands derived by \texttt{MUFFIT}. These parameters were estimated assuming \citet[][BC03]{bc03} stellar population models, \citet{fitzpatrick99} extinction law, and \citet{chabrier03} initial mass function (IMF). We refer the reader to \citet{diazgarcia15}, \citet{diazgarcia18sp} and DG17 for further details about \texttt{MUFFIT} and derived quantities.

\subsection{Selection of quiescent and star-forming galaxies}\label{selection}
Throughout this paper, we focus our analysis on the galaxies in the ALHAMBRA gold catalogue\footnote{\tt http://cosmo.iaa.es/content/ALHAMBRA-Gold-catalog}. This catalogue comprises $\sim100$k sources with $F814W \leq 23$ \citep{molino13}.

We split our galaxies into quiescent and star-forming with the dust-corrected version of the $UVJ$ colour-colour plane selection presented in DG17, adapted to the ALHAMBRA medium-band filter system: we used $F365$ instead of the filter $U$ and $F551$ instead of the filter $V$. The ALHAMBRA filter $J$ is the standard one. As shown by DG17, quiescent and star-forming galaxies with $F814W \leq 23$ define two non-overlapping populations in the colour-colour plane after removing dust effects, with the selection boundary located at $(F365 - F551) = 1.5$. We refer the reader to DG17 for a detailed description of the selection process and the study of the stellar population properties of quiescent galaxies in the $UVJ$ colour-colour plane. We show the observed (i.e. reddened by dust) rest-frame distribution of the 76642 ALHAMBRA gold catalogue galaxies with $z < 1.5$ in Fig.~\ref{UVJ}. The quiescent population is enclosed by the common colour-colour selection box \citep{williams09}, but a population of dusty star-forming galaxies is also located in this area. DG17 show that a significant fraction ($\sim 20$\%) of the red galaxies are indeed dusty star-forming, contaminating the quiescent population. Thanks to the low-resolution spectral information from ALHAMBRA, the \texttt{MUFFIT} code is able to provide a robust quiescent vs. star-forming classification.

The final sample, located at $z < 1.5$ with $F814W \leq 23$, comprises 12905 quiescent and 63737 star-forming galaxies. The stellar masses covered by our data span the $8 < \log_{10} M_{\star} / M_{\odot} < 11.5$ range. Further details about the stellar mass completeness and the redshift distribution of the sample are presented in DG17. We study the MLCR of these samples in the next section.

\section{Mass-to-light ratio vs. colour relation at $z < 1.5$}\label{mlratio}
In this section, we study the relation between the mass-to-light ratio in the $i$ band and the observed rest-frame $(g-i)$ colour of the ALHAMBRA galaxies with $z < 1.5$. In some cases, we denote $\Upsilon = \log_{10}(M_{\star}/L_i)$ and ${\mathcal C} = (g-i)$ for the sake of clarity. The redshift, stellar masses, and observed rest-frame (i.e. reddened by dust) luminosities were derived by the \texttt{MUFFIT} code (Sect.~\ref{muffit}).

\begin{figure}[t]
\centering
\resizebox{\hsize}{!}{\includegraphics{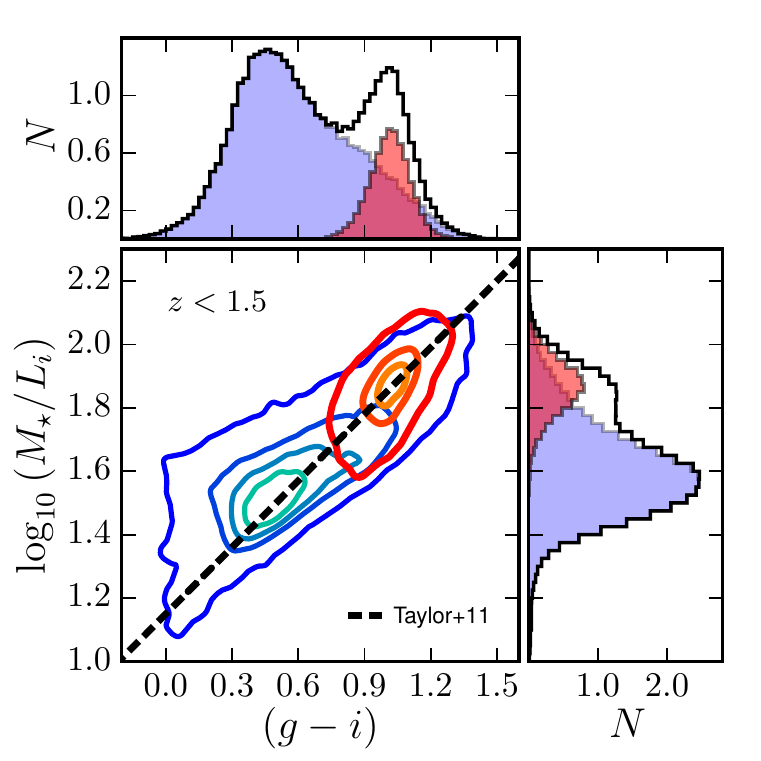}}\\
\resizebox{\hsize}{!}{\includegraphics{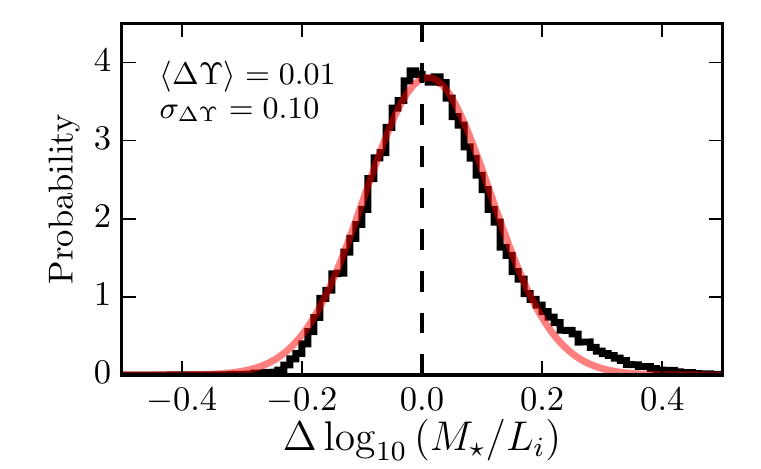}}
\caption{{\it Top panel}: Mass-to-light ratio $M_{\star}/L_{i}$ as a function of the rest-frame colour $g-i$ for the 76642 ALHAMBRA galaxies with $F814W \leq 23$ at $z < 1.5$. The coloured lines show level contours in the density of galaxies, starting at 0.1 galaxies dex$^{-2}$ and increasing in 1.5 galaxies dex$^{-2}$ steps. Red contours show the quiescent population, and blue contours the star-forming one. The side panels show the normalized distribution in $g-i$ ({\it upper panel}) and $\log_{10}\,(M_{\star}/L_{i})$ ({\it right panel}). In both panels the total distribution is presented in black, the quiescent population in filled red, and the star-forming population in filled blue. The black dashed line marks the relation derived by \citet{taylor11} at $z < 0.65$ using GAMA galaxies and SDSS five-band photometry. {\it Bottom panel}: Comparison between the observed ALHAMBRA mass-to-light ratio and that expected from the \citet{taylor11} relation. The red solid line is the best Gaussian fit with median $\langle \Delta \Upsilon \rangle = 0.01$ dex and dispersion $\sigma_{\Delta \Upsilon} = 0.10$ dex.}
\label{MLtot}
\end{figure}

We present the $M_{\star}/L_{i}$ vs. $(g-i)$ colour plane for both quiescent and star-forming galaxies in the {\it top panel} of Fig.~\ref{MLtot}. We find that, for both populations, the mass-to-light ratio increases for redder colours, in agreement with the literature (see references in Sect.~\ref{intro}). We describe the modelling of this dependence in Sect.~\ref{modelling}. In the figure, we also present the MLCR found by T11 in the GAMA survey. Their relation is in excellent agreement with our observed values: the comparison between our measurements and their predictions, $\Delta \Upsilon = \Upsilon - \Upsilon_{\rm T11}({\mathcal C})$, has no bias, $\langle \Delta \Upsilon \rangle = 0.01$ dex, and a small dispersion of $\sigma_{\Delta \Upsilon} = 0.1$ dex ({\it bottom panel} in Fig.~\ref{MLtot}), similar to the one found by T11 with GAMA data. We note that T11 use BC03 stellar population models and a \citet{chabrier03} IMF, as we did, but different extinction laws (\citealt{calzetti00} vs. \citealt{fitzpatrick99}) and star formation histories (SFHs; $e-$fold tau models vs. two stellar populations mix) were assumed.

The T11 study is performed in a sample of $z < 0.65$ galaxies with a median redshift of $\langle z \rangle = 0.2$, and our data covers a wider redshift range ($z \leq 1.5$) with a median redshift of $\langle z \rangle = 0.65$. This suggests that the low-redshift relation measured by T11 in GAMA has not evolved significantly with redshift. We assume this redshift independence in the following and test it in Sect.~\ref{MLz}.

\subsection{Modelling the intrinsic mass-to-light vs. colour relation}\label{modelling}
The measurements presented in the previous section are affected by observational errors, blurring the information and biasing our analysis. We are interested in the intrinsic distribution of our measurements in the mass-to-light ratio vs. colour plane, and in this section we detail the steps to estimate it. The results are presented in Sect.~\ref{results}.

The intrinsic distribution of interest is noted $D$, and provides the real values of our measurements for a set of parameters $\theta$,
\begin{equation}
D\,(\Upsilon_0, {\mathcal C}_0\,|\,\theta),
\end{equation}
where $\Upsilon_0$ and ${\mathcal C}_0$ are the real values of the mass-to-light ratio and the colour unaffected by observational errors. We derive the posterior of the parameters $\theta$ that define the intrinsic distribution $D$ for both quiescent and star-forming galaxies with a Bayesian model. Formally,
\begin{equation}
P\,(\theta\,|\,\Upsilon,{\mathcal C},\sigma_{\Upsilon},\sigma_{\mathcal C}) \propto {\mathcal L}\,(\Upsilon,{\mathcal C}\,|\,\theta, \sigma_{\Upsilon}, \sigma_{\mathcal C})\,P(\theta),
\end{equation}
where $\sigma_{\Upsilon}$ and $\sigma_{\mathcal C}$ are the uncertainties in the observed mass-to-light ratio and $(g-i)$ colour, respectively, ${\mathcal L}$ is the likelihood of the data given $\theta$, and $P(\theta)$ the prior in the parameters. The posterior probability is normalised to one.

The likelihood function associated to our problem is
\begin{equation}
{\mathcal L}\,(\Upsilon,{\mathcal C}\,|\,\theta, \sigma_{\Upsilon}, \sigma_{\mathcal C}) = \prod_k P_k\,(\Upsilon_k,{\mathcal C}_k\,|\,\theta, \sigma_{\Upsilon,k}, \sigma_{{\mathcal C},k}),
\end{equation}
where the index $k$ spans the galaxies in the sample, and $P_k$ traces the probability of the measurement $k$ for a set of parameters $\theta$. This probability can be expressed as
\begin{align}
P_k\,(&\Upsilon_k,{\mathcal C}_k\,|\,\theta, \sigma_{\Upsilon,k}, \sigma_{{\mathcal C},k}) = \nonumber\\ 
&\int \! D\,(\Upsilon_0, {\mathcal C}_0\,|\,\theta)\,P_G(\Upsilon_k\,|\,\Upsilon_0, \sigma_{\Upsilon,k})\,P_G({\mathcal C}_k\,|\,{\mathcal C}_0, \sigma_{{\mathcal C},k})\,{\rm d}\Upsilon_0\,{\rm d}{\mathcal C}_0,\label{eq_pk}
\end{align}
where the real values $\Upsilon_0$ and ${\mathcal C}_0$ derived from the model $D$ are affected by Gaussian observational errors,
\begin{equation}
P_G\,(x\,|\,x_0, \sigma_x) = \frac{1}{\sqrt{2 \pi} \sigma} \exp\bigg[-\frac{(x - x_0)^2}{2\sigma^2}\bigg],
\end{equation}
providing the likelihood of observing a magnitude given its real value and uncertainty. We have no access to the real values $\Upsilon_0$ and ${\mathcal C}_0$, so we marginalise over them in Eq.~(\ref{eq_pk}) and the likelihood is expressed therefore with known quantities. We assumed no covariance between $\Upsilon$ and ${\mathcal C}$, although they share the $i$-band luminosity information. We checked by Monte Carlo sampling of the $M_{\star}$, $L_i$, and $L_g$ distributions that such covariance is small, with $\rho_{\Upsilon{\mathcal C}} \sim 0.05$. Hence, we disregard the covariance term by simplicity.

We explore the parameters posterior distribution with the \texttt{emcee} \citep{emcee} code, a \texttt{Python} implementation of the affine-invariant ensemble sampler for Markov chain Monte Carlo (MCMC) proposed by \citet{goodman10}. The \texttt{emcee} code provides a collection of solutions in the parameter space, noted $\theta_{\rm MC}$, with the density of solutions being proportional to the posterior probability of the parameters. We obtained central values of the parameters as the median, noted $\langle \theta_{\rm MC} \rangle$, and their uncertainties as the range enclosing 68\% of the projected solutions around the median.

We define in the following the distributions assumed for the quiescent and star-forming populations, and the prior imposed to their parameters. The quiescent (${\rm Q}$) population is described as
\begin{equation}
D_{\rm Q}\,(\Upsilon_0, {\mathcal C}_0\,|\,\theta_{\rm Q}) = P_G\,({\mathcal C}_0\,|\,\mu_{\rm Q}, s_{\rm Q})\,P_G\,(\Upsilon_0\,|\,A_{\rm Q} + B_{\rm Q}\,{\mathcal C}_0, \sigma_{\rm Q}),
\end{equation}
where $\mu_{\rm Q}$ and $s_{\rm Q}$ describe the intrinsic $(g-i)$ colour distribution, $A_{\rm Q}$ and $B_{\rm Q}$ are the coefficients that define the MLCR, and $\sigma_{\rm Q}$ the intrinsic (i.e. related to physical processes) dispersion of such relation. We have a set of five parameters to describe the distribution of quiescent galaxies, $\theta_{\rm Q} = \{\mu_{\rm Q}, s_{\rm Q}, A_{\rm Q}, B_{\rm Q}, \sigma_{\rm Q}\}$. We used flat priors, $P(\theta_{\rm Q}) = 1$, except for the dispersions $s_{\rm Q}$ and $\sigma_{\rm Q}$, that we imposed as positive.

\begin{figure}[t]
\centering
\resizebox{\hsize}{!}{\includegraphics{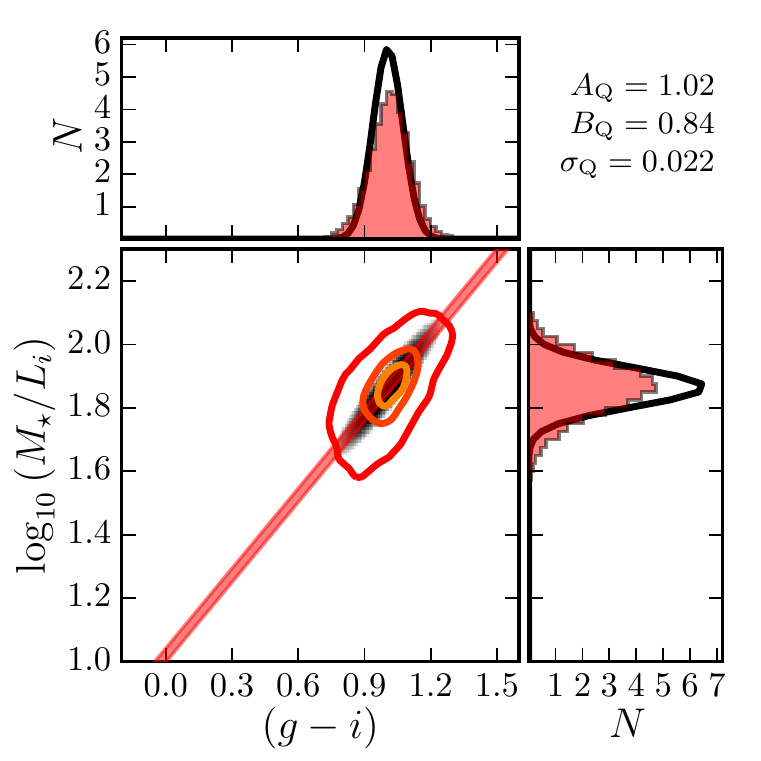}}\\
\resizebox{\hsize}{!}{\includegraphics{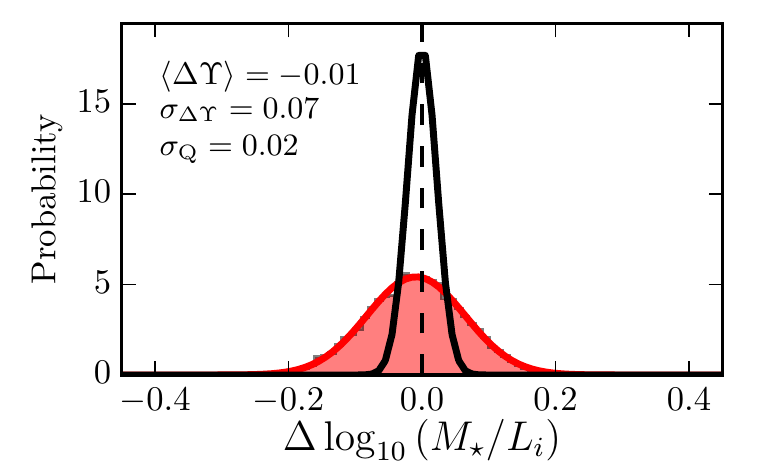}}
\caption{{\it Top panel}: Mass-to-light ratio $M_{\star}/L_{i}$ as a function of the rest-frame colour $g-i$ for the 12905 ALHAMBRA quiescent galaxies with $F814W \leq 23$ at $z < 1.5$. The solid lines show level contours in the density of galaxies as in Fig.~\ref{MLtot}. The grey scale shows the median fitting model to the data, $D_{\rm Q}\,(\Upsilon_0, \mathcal{C}_0\,|\,\langle \theta_{\rm Q} \rangle)$. The red area represents the derived MLCR, $\log_{10}\,(M_{\star}/L_{i}) = 1.02 + 0.84(g-i)$, and its $1\sigma$ intrinsic dispersion, $\sigma_{\rm P} = 0.02$. The side panels show the normalized projected histogram in $g-i$ ({\it upper panel}) and $\log_{10}\,(M_{\star}/L_{i})$ ({\it right panel}). In both panels the observed distribution is presented in filled red, and the derived median model in solid black. {\it Bottom panel}: Comparison between the observed ALHAMBRA mass-to-light ratio and the expected from our median relation (red filled histogram). The solid red line is the best Gaussian fit with median $\langle \Delta \Upsilon \rangle = -0.01$ and $\sigma_{\Delta \Upsilon} = 0.07$. The solid black line illustrates the estimated intrinsic dispersion unaffected by observational uncertainties.}
\label{MLred}
\end{figure}

\begin{figure}[t]
\centering
\resizebox{\hsize}{!}{\includegraphics{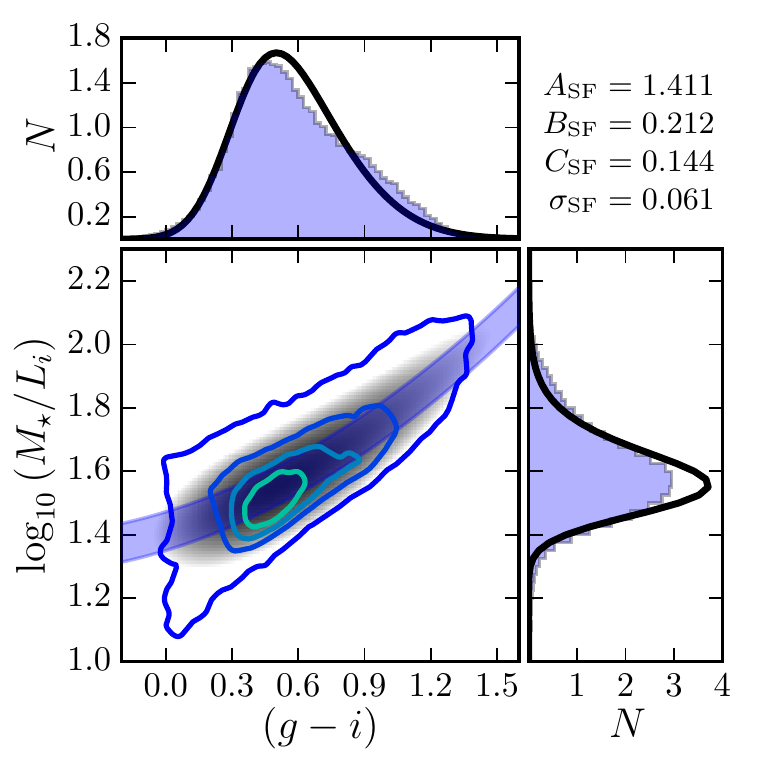}}\\
\resizebox{\hsize}{!}{\includegraphics{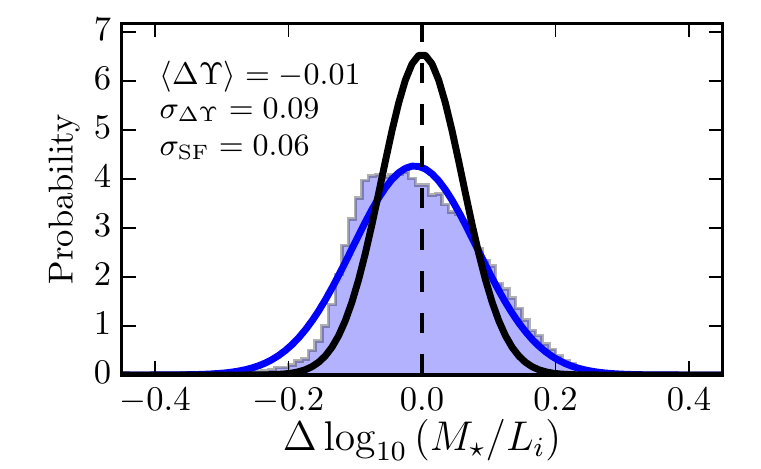}}
\caption{{\it Top panel}: Mass-to-light ratio $M_{\star}/L_{i}$ as a function of the rest-frame colour $g-i$ for the 63737 ALHAMBRA star-forming galaxies with $F814W \leq 23$ at $z < 1.5$. The solid lines show level contours in the density of galaxies as in Fig.~\ref{MLtot}. The grey scale shows the median fitting model to the data, $D_{\rm SF}\,(\Upsilon_0, \mathcal{C}_0\,|\,\langle \theta_{\rm SF} \rangle)$. The blue area represents the derived MLCR, $\log_{10}\,(M_{\star}/L_{i}) = 1.411 + 0.212(g-i) + 0.144(g-i)^2$, and its $1\sigma$ intrinsic dispersion, $\sigma_{\rm SF} = 0.06$. The side panels show the normalized projected histogram in $g-i$ ({\it upper panel}) and $\log_{10}\,(M_{\star}/L_{i})$ ({\it right panel}). In both panels the observed distribution is presented in filled blue, and the derived median model in solid black. {\it Bottom panel}: Comparison between the observed ALHAMBRA mass-to-light ratio and the expected from our median relation (blue filled histogram). The blue solid line is the best Gaussian fit with median $\langle \Delta \Upsilon \rangle = -0.01$ and $\sigma_{\Delta \Upsilon} = 0.09$. The black solid line illustrates the estimated intrinsic dispersion unaffected by observational uncertainties.}
\label{MLblue}
\end{figure}

The star-forming (${\rm SF}$) population presents a more complex behaviour (Fig.~\ref{MLtot}), and we modelled it as
\begin{align}
D_{\rm SF}\,(\Upsilon_0, {\mathcal C}_0\,|\,\theta_{\rm SF}) =&\,P_G\,({\mathcal C}_0\,|\,\mu_{\rm SF}, s_{\rm SF})\,\bigg[1 + {\rm erf}\,\bigg(\alpha_{\rm SF}\,\frac{C_0 - \mu_{\rm SF}}{\sqrt{2}s_{\rm SF}}\bigg)\bigg]\nonumber\\
&P_G\,(\Upsilon_0\,|\,A_{\rm SF} + B_{\rm SF}\,{\mathcal C}_0 + C_{\rm SF}\,{\mathcal C}_0^2, \sigma_{\rm SF}),
\end{align}
where $\mu_{\rm SF}$, $s_{\rm SF}$, and $\alpha_{\rm SF}$ describe the intrinsic $(g-i)$ colour distribution, $A_{\rm SF}$, $B_{\rm SF}$ and $C_{\rm SF}$ are the coefficients that define the MLCR for star-forming galaxies, and $\sigma_{\rm SF}$ the intrinsic dispersion of such relation. Important differences are present as compared with the quiescent population. First, the distribution of ${\mathcal C}_0$ is not symmetric (Fig.~\ref{MLtot}). We accounted for this asymmetry by adding the error function term and the parameter $\alpha_{\rm SF}$, that controls the skewness of the distribution \citep{azzalini05}. Second, we found that the dependence of $\Upsilon_0$ with the colour is not linear, but a second order polynomial. This is motivated by the apparent curvature present at the redder colours in Fig.~\ref{MLtot}. To choose between the linear or the parabolic MLCR, we used the Bayesian information criterion (BIC, \citealt{schwarz78}), defined as
\begin{equation}
{\rm BIC} = N\log n - 2\log {\mathcal L}\,(\Upsilon, {\mathcal C}\,|\,\langle \theta_{\rm MC} \rangle),
\end{equation}
where $N$ is the number of parameters in the model and $n$ the number of galaxies in the sample. We find $\Delta {\rm BIC} = {\rm BIC}_{\rm par} - {\rm BIC}_{\rm lin} = -750$, favouring the inclusion of $C_{\rm SF}$ in the modelling. For consistency, we checked the application of a parabolic MLCR for quiescent galaxies. We found $\Delta {\rm BIC} = {\rm BIC}_{\rm par} - {\rm BIC}_{\rm lin} = 1.5$, thus favouring the simpler linear model. Figure~\ref{MLtot} also suggests an asymmetric distribution in $\Upsilon_0$, instead of the assumed Gaussian. We studied the inclusion of an additional skew parameter for $\Upsilon_0$, but it was consistent with zero and in this case the BIC favours the simpler Gaussian model without the extra skew parameter. Finally, we have a set of seven parameters to describe the distribution of star-forming galaxies, $\theta_{\rm SF} = \{\mu_{\rm SF}, s_{\rm SF}, \alpha_{\rm SF}, A_{\rm SF}, B_{\rm SF}, C_{\rm SF}, \sigma_{\rm SF}\}$. We used flat priors, $P(\theta_{\rm SF}) = 1$, except for the dispersions $s_{\rm SF}$ and $\sigma_{\rm SF}$, that we imposed as positive.

We note that the redshift dimension is not included in our analysis because we are assuming that the MLCRs do not depend on redshift. This was initially motivated by the excellent agreement with the local relation from T11 shown in Fig.~\ref{MLtot}, and we further test this assumption in Sect.~\ref{MLz}.

\section{Results}\label{results}
We present the derived $i-$band MLCRs for both quiescent and star-forming galaxies in Sect.~\ref{mlratiomodel}, and explore the redshift dependence of the relations in Sect.~\ref{MLz}. The $g-$ and $r-$band MLCRs are presented in Sect.~\ref{MLgr}, and we compare our results with the literature in Sect.~\ref{mlratiolit}.

\subsection{Mass-to-light ratio vs. colour relation for quiescent and star-forming galaxies}\label{mlratiomodel}
In this section, we present the results of our modelling. They are summarised on Fig.~\ref{MLred} for quiescent galaxies and Fig.~\ref{MLblue} for the star-forming ones. The derived parameters are compiled in Table~\ref{param_tab}. We find that, in both cases, the assumed model describes satisfactorily the observed distributions in colour and mass-to-light ratio spaces. 

We start presenting the results for quiescent galaxies. We estimate
\begin{equation}
\Upsilon_{\rm Q} = 1.02 + 0.84\,(g-i)\label{eq_mlred}
\end{equation}
with a small intrinsic dispersion of $\sigma_{\rm Q} = 0.02$ dex. The observed dispersion, that includes the observational errors, was estimated from a Gaussian fit to the distribution of the variable $\Delta \Upsilon = \Upsilon - \Upsilon_{\rm Q}$, yielding $\sigma_{\Delta \Upsilon} = 0.07$ dex ({\it bottom panel} in Fig.~\ref{MLred}). This value is lower than the 0.1 dex obtained with the local MLCR from T11.

\begin{figure}[t]
\centering
\resizebox{\hsize}{!}{\includegraphics{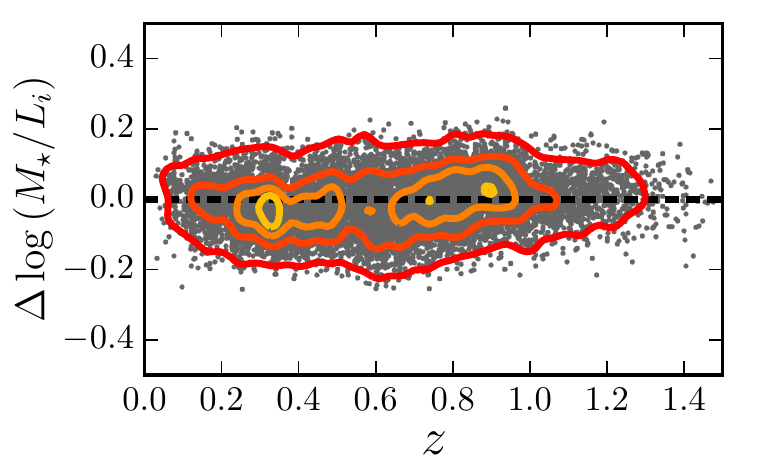}}\\
\resizebox{\hsize}{!}{\includegraphics{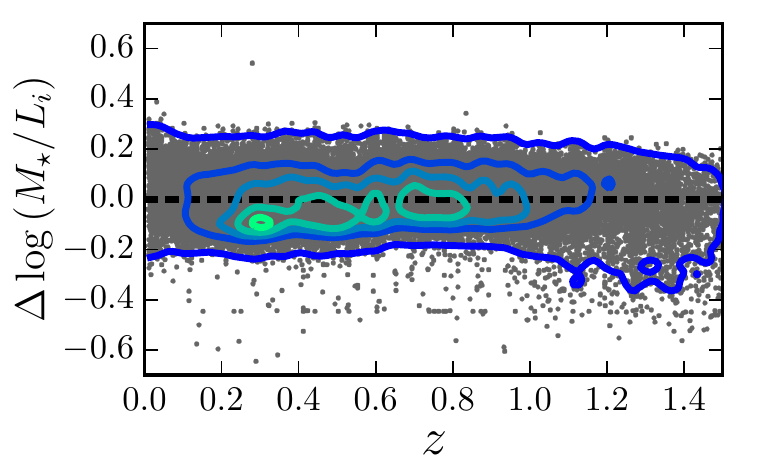}}
\caption{Comparison between the observed ALHAMBRA mass-to-light ratio and the expected one from our median relation as a function of redshift for quiescent ({\it top panel}) and star-forming ({\it bottom panel}) galaxies with $F814W \leq 23$ (gray dots). The solid lines show level contours in the number of galaxies, starting at one galaxy and increasing in five galaxies steps for quiescent galaxies, and in 15 galaxies steps for star-forming galaxies. The dashed lines mark null difference.}
\label{DMLz}
\end{figure}

For star-forming galaxies, we find
\begin{equation}
\Upsilon_{\rm SF} = 1.411 + 0.212\,(g-i) + 0.144\,(g-i)^2\label{eq_mlblue}
\end{equation}
with an intrinsic dispersion of $\sigma_{\rm SF} = 0.06$ dex. The observed dispersion in this case is $\sigma_{\Delta \Upsilon} = 0.09$ dex ({\it bottom panel} in Fig.~\ref{MLblue}), similar to the 0.1 dex obtained with the T11 relation. The higher complexity of the star-forming population is not surprising due to the combination of an underlying old population that dominates the stellar mass, a young population dominating the emission in the bluer bands, and the presence of different dust contents. Despite this fact, a well defined MLCR with a small dispersion is inferred from our data.

We conclude that the encouraging 0.1 dex precision in the mass-to-light ratio estimation from the optical colour $(g-i)$ found by T11 is even tighter after the observational uncertainties are accounted for. The dispersion derived with ALHAMBRA data at $z < 1.5$ is 0.02 dex for quiescent galaxies and 0.06 dex for star-forming galaxies. These small dispersions refer to the statistical analysis of the data, and systematic uncertainties related with the assumed stellar population models, IMF, SFHs, extinction law, etc. are not included in the analysis (see \citealt{portinari04}, \citealt{barro11mass}, and \citealt{courteau14}, for a detailed discussion about systematics in stellar mass estimations). The similarity between T11 and our values suggests that the assumed extinction law and the SFHs are not an important source of systematics, with stellar population models and the IMF being the main contributors. The application of different stellar population models, such as those from \citet{vazdekis16}, \citet{maraston05}, or \citet{conroy10}, is beyond the scope of the present work.

\begin{table*}[t]
\caption{ALHAMBRA mass-to-light ratio vs. $(g-i)$ colour relation.}
\label{param_tab}
\begin{center}
\begin{tabular}{lcccccc}
\hline\hline\noalign{\smallskip}
Optical band & Galaxy type &  $A$ & $B$ & $C$ & $\sigma_{\rm int}$ & $\sigma_{\Delta \Upsilon}$\\
\noalign{\smallskip}
\hline
\noalign{\smallskip}
$i$ band & Quiescent	& $1.02 \pm 0.01$ & $0.84 \pm 0.01$ &          $\cdots$ & $0.022 \pm 0.001$ & $0.07$\\
         & Star-forming	& $1.411 \pm 0.003$ & $0.212 \pm 0.007$ & $0.144 \pm 0.005$ & $0.061 \pm 0.001$ & $0.09$\\
\noalign{\smallskip}
\hline
\noalign{\smallskip}
$r$ band & Quiescent	& $1.02 \pm 0.01$ & $1.01 \pm 0.01$ &          $\cdots$ & $0.021 \pm 0.001$ & $0.08$\\
         & Star-forming	&  $1.453 \pm 0.003$ & $0.373 \pm 0.008$ & $0.128 \pm 0.007$ & $0.063 \pm 0.001$ & $0.10$\\
\noalign{\smallskip}
\hline
\noalign{\smallskip}
$g$ band & Quiescent	& $0.98 \pm 0.02$ & $1.28 \pm 0.02$ &          $\cdots$ & $0.014 \pm 0.001$ & $0.07$\\
         & Star-forming	& $1.386 \pm 0.003$ & $0.707 \pm 0.009$ & $0.078 \pm 0.007$ & $0.057 \pm 0.001$ & $0.09$\\
\hline

\end{tabular}
\end{center}
\end{table*}

\subsection{Redshift evolution of the mass-to-light ratio vs. colour relation}\label{MLz}
The results presented in previous section implies a tight relation of the mass-to-light ratio with the optical colour $(g-i)$. In our analysis, we assumed such a relation as redshift independent, motivated by the nice agreement with the $z \sim 0.2$ results from T11 (Fig.~\ref{MLtot}).

We present the redshift evolution of $\Delta \Upsilon$ in Fig.~\ref{DMLz}, both for quiescent and star-forming galaxies. We find no evidence of redshift evolution either for quiescent or star-forming galaxies. The median $|\Delta \Upsilon|$ at any redshift is always below 0.02 dex, and a simple linear fitting constrains the possible residual evolution with $z$ to less than $0.05$ dex since $z = 1.5$. We conclude therefore that the relations presented in Eq.~(\ref{eq_mlred}) and Eq.~(\ref{eq_mlblue}) have not changed appreciably during the last 9 Gyr of the Universe, with quiescent and star-forming galaxies galaxies evolving along the derived relations since $z = 1.5$.


\subsection{Mass-to-light ratio vs. colour relation in the $r$ and $g$ bands}\label{MLgr}
We complement the results in the previous sections with the estimation of the intrinsic relation between the mass-to-light ratio in the $r$ and $g$ bands with $(g-i)$ colour, both for quiescent and star-forming galaxies. We confirm the tight relations found with the $i$-band luminosity and the curvature for the star-forming population. We present the estimated relations in Table~\ref{param_tab} for future reference.

We find that the normalization of the MLCRs are similar in the $gri$ bands at 0.05 dex level. This is because our luminosities are expressed in AB units, so a null colour implies the same luminosity in all the bands, which share a common stellar mass.

Regarding the slope for the quiescent population, it is larger for bluer bands. This implies that at the median colour of the quiescent population, $\langle (g-i) \rangle = 1$, the mass-to-light ratio decreases from $\log_{10}(M_{\star}/L_{g}) = 2.26$ to $\log_{10}(M_{\star}/L_{i}) = 1.86$, reflecting the larger contribution to the stellar mass budget of redder low-mass stars.

In the case of the star-forming galaxies, the parameter $B_{\rm SF}$ is larger at bluer bands, but the parameter $C_{\rm SF}$ is smaller. This implies a lower curvature of the MLCR in the $g$ band. We checked that the quadratic model is still favoured by the data even in the $g$-band case.

The intrinsic dispersion in the MLCRs is still low and similar to the $i$-band values, with $\sigma_{\rm P} \sim 0.02$ dex and $\sigma_{\rm SF} \sim 0.06$ dex. Finally, the observed dispersion, affected by observational errors, are also similar to the fiducial $i$-band values, as summarised in Table~\ref{param_tab}.

We conclude that the MLCR holds in the optical range covered by the $gri$ bands, confirming the tight correlation between optical mass-to-light ratios and the rest-frame colour $(g-i)$. 

\begin{figure}[t]
\centering
\resizebox{\hsize}{!}{\includegraphics{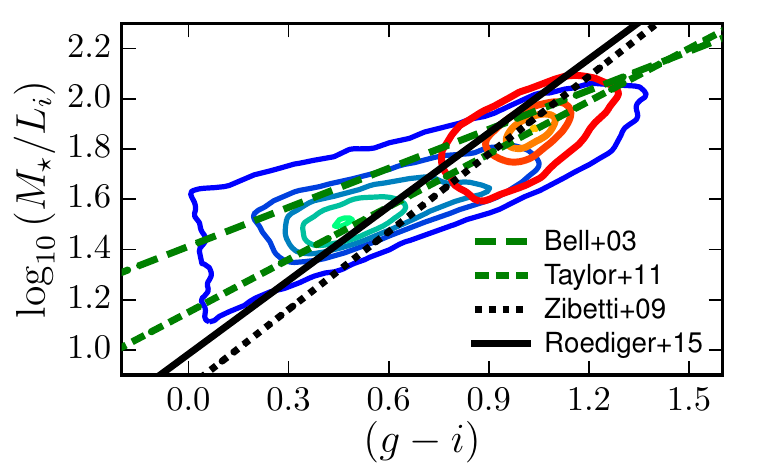}}
\caption{Comparison of the observed mass-to-light ratio vs. $(g-i)$ colour in ALHAMBRA with MLCRs from the literature. The contours are the same as in Fig.~\ref{MLtot}. The dashed green line is from the observational study of T11. The other black lines are from theoretical expectations: \citet[][dotted]{zibetti09} and \citet[][solid]{roediger15}. All the MLCRs have been scaled to a \citet{chabrier03} IMF and referred to BC03 models.}
\label{MLlit}
\end{figure}

\subsection{Comparison with the literature}\label{mlratiolit}
In addition to the T11 work, several studies in the literature have tackled the problem of the MLCR, both theoretically and observationally (see references in Sect.~\ref{intro}). We present the $i$-band mass-to-light ratio vs. $(g-i)$ colour from previous work in Fig.~\ref{MLlit}. We only present the colour range imposed by the ALHAMBRA data, $0 < (g-i) < 1.5$ (Fig.~\ref{MLtot}). All the MLCRs have been scaled to a \citet{chabrier03} IMF and referred to BC03 stellar population models to minimise systematic differences.

We find a reasonably good agreement with the theoretical results from \citet{roediger15} and \citet{zibetti09}. The comparison of these predictions with our values yields a bias of $\langle \Delta \Upsilon \rangle = -0.01$ and $0.08$, and a dispersion of $\sigma_{\Delta \Upsilon} = 0.17$ and $0.19$, respectively. We highlight the predictions from \citet{roediger15}, that have no bias and only a factor of two larger dispersion than our optimal MLCRs.

From the observational point of view, we recall the agreement with the results from T11 (Fig.~\ref{MLtot}). Their relation provides no bias and a dispersion of $\sigma_{\Delta \Upsilon} = 0.1$. We also compare our results with the popular work by \citet{bell03}. They relation yields a bias of $\langle \Delta \Upsilon \rangle = -0.12$ and again a dispersion of $\sigma_{\Delta \Upsilon} = 0.1$. We note that the MLCR of \citet{bell03} was estimated with \texttt{PEGASE} \citep{pegase} stellar populations models and a ``diet Salpeter'' IMF. Hence, we applied to the relations in \citet{bell03} a -0.10 dex offset to account for the differentce in the stellar population models, as estimated by \citet{barro11mass}, and a -0.15 dex offset to scale the IMF. 

Following T11, we conclude that the range of colours covered by the observed galaxies, that are consequence of their formation and evolution, restrict the parameter space of the models and provide tighter MLCRs than expected from theory. The bias with respect to previous work is at $\sim 0.1$ dex level, supponting the tight relations derived from ALHAMBRA data.

\section{Summary and conclusions}\label{conclusions}
We used the redshifts, stellar masses and rest-frame colours derived with \texttt{MUFFIT} for 76642 ALHAMBRA sources at $z \leq 1.5$ to explore the $i$-band mass-to-light ratio relation with the rest-frame $(g-i)$ colour. As shown by T11, there is a tight (0.1 dex) MLCR in the GAMA survey at $z \sim 0.2$, and we expand their study up to $z = 1.5$.

We found that the $i-$band MLCR is also present in ALHAMBRA at $z \leq 1.5$, for both quiescent and star-forming galaxies. The data suggests a lineal MLCR for quiescent galaxies and a quadratic one for star-forming systems, as summarised in Table~\ref{param_tab}, and also holds for $g$ and $r$ luminosities. These relations present an intrinsic dispersion, after accounting by observational uncertainties, of $\sigma_{\rm P} = 0.02$ dex and $\sigma_{\rm SF} = 0.06$ dex. These dispersions are intrinsic, and must be accounted in addition to the observational uncertainties of the colour. We also stress that they refer to statistical dispersions, and the final error budget in mass-to-light ratio predictions should account by systematic uncertainties ($\sim0.2$ dex; e.g. \citealt{barro11mass}) related with the assumed stellar population models, IMF, SFHs, extinction law, etc.

Our measurements suggests that the estimated MLCRs are redshift-independent at least since $z \sim 1.5$. This is, quiescent and star-forming galaxies have evolved along the MLCRs in the last 9 Gyrs of the Universe, preserving the observed relations with time.

We compare our data with other proposed MLCRs in the literature. The observational relation of T11, based on GAMA survey data, reproduces our values with no bias and dispersion $\sigma_{\Delta \Upsilon} = 0.1$ dex. Regarding theoretical studies, the MLCR from \citet{roediger15} matches best with our measurements, the bias is below 0.1 dex and the dispersion is $\sigma_{\Delta \Upsilon} = 0.17$ dex.

Our results could be expanded in several ways. The analysis could be made by using different stellar population models to test the redshift independence of the relations and the curvature of the star-forming MLCR. The study of the MLCR at higher redshifts will provide extra clues about the absence of redshift evolution, for which a NIR-selected ALHAMBRA sample is needed \citep{nieves17}. Finally, the study at masses lower than $\log_{10} M_{\star} \sim 8$ will test the results' robustness at the bluer end of the relation, where intense star-forming episodes could compromise the stellar masses estimated with our current techniques. 

The derived relations can be used to estimate stellar masses with photometric redshift codes based on a limited set of empirical templates, such as \texttt{BPZ2}. The intrinsic MLCRs, unaffected by observational errors, are the needed priors to define the probability distribution function (PDF) of the stellar mass. The PDF-based estimator of the luminosity function was presented by \citet{clsj17lfbal} as part of the PROFUSE\footnote{\tt profuse.cefca.es} project, that uses PRObability Functions for Unbiased Statistical Estimations in multi-filter surveys, and successfully applied to estimate the $B$-band luminosity function at $z < 1$ \citep{clsj17lfbal} and the $UV$ luminosity function at $2.5 \leq z < 4.5$ \citep{viironen18} in ALHAMBRA. The present paper is a fundamental step towards a PDF-based estimator of the stellar mass function.

\begin{acknowledgements}
We dedicate this paper to the memory of our six IAC colleagues and friends who
met with a fatal accident in Piedra de los Cochinos, Tenerife, in February 2007,
with a special thanks to Maurizio Panniello, whose teachings of \texttt{python}
were so important for this paper.

We thank R. Angulo, S. Bonoli, A. Ederoclite, C. Hern\'andez-Monteagudo, A. Mar\'{\i}n-Franch, A. Orsi, and all the CEFCA staff, post-docs, and students for useful and productive discussions.

This work has been mainly funding by the FITE (Fondos de Inversiones de Teruel) and the Spanish MINECO/FEDER projects AYA2015-66211-C2-1-P, AYA2012-30789, AYA2006-14056, and CSD2007-00060. We also acknowledge the financial support from the Arag\'on Government Research Group E96 and E103.

We acknowledge support from the Spanish Ministry for Economy and Competitiveness and FEDER funds through grants AYA2010-15081, AYA2010-22111-C03-01, AYA2010-22111-C03-02, AYA2012-39620, AYA2013-40609-P, AYA2013-42227-P, AYA2013-48623-C2-1, AYA2013-48623-C2-2, AYA2016-76682-C3-1-P, AYA2016-76682-C3-3-P, ESP2013-48274, Generalitat Valenciana project Prometeo PROMETEOII/2014/060, Junta de Andaluc\'{\i}a grants TIC114, JA2828, P10-FQM-6444, and Generalitat de Catalunya project SGR-1398.

K. V. acknowledges the {\it Juan de la Cierva incorporaci\'on} fellowship, IJCI-2014-21960, of the Spanish government. A. M. acknowledges the financial support of the Brazilian funding agency FAPESP (Post-doc fellowship - process number 2014/11806-9). B. A. has received funding from the European Union’s Horizon 2020 research and innovation programme under the Marie Sklodowska-Curie grant agreement No. 656354. M. P. acknowledges financial supports from the Ethiopian Space Science and Technology Institute (ESSTI) under the Ethiopian Ministry of Science Science and Technology (MoST).

This research made use of \texttt{Astropy}, a community-developed core \texttt{Python} package for Astronomy \citep{astropy}, and \texttt{Matplotlib}, a 2D graphics package used for \texttt{Python} for publication-quality image generation across user interfaces and operating systems \citep{pylab}.

\end{acknowledgements}

\bibliography{biblio}

\begin{thebibliography}{63}
\expandafter\ifx\csname natexlab\endcsname\relax\def\natexlab#1{#1}\fi

\bibitem[{{Abazajian} {et~al.}(2009){Abazajian}, {Adelman-McCarthy},
  {Ag{\"u}eros}, {Allam}, {Allende Prieto}, {An}, {Anderson}, {Anderson},
  {Annis}, {Bahcall}, \& et~al.}]{sdssdr7}
{Abazajian}, K.~N., {Adelman-McCarthy}, J.~K., {Ag{\"u}eros}, M.~A., {et~al.}
  2009, \apjs, 182, 543

\bibitem[{{Aparicio-Villegas} {et~al.}(2010){Aparicio-Villegas}, {Alfaro},
  {Cabrera-Ca{\~n}o}, {Moles}, {Ben{\'{\i}}tez}, {Perea}, {del Olmo},
  {Fern{\'a}ndez-Soto}, {Crist{\'o}bal-Hornillos}, {Husillos}, {Aguerri},
  {Broadhurst}, {Castander}, {Cepa}, {Cervi{\~n}o}, {Gonz{\'a}lez Delgado},
  {Infante}, {M{\'a}rquez}, {Masegosa}, {Mart{\'{\i}}nez}, {Prada}, {Quintana},
  \& {S{\'a}nchez}}]{aparicio10}
{Aparicio-Villegas}, T., {Alfaro}, E.~J., {Cabrera-Ca{\~n}o}, J., {et~al.}
  2010, \aj, 139, 1242

\bibitem[{{Arnalte-Mur} {et~al.}(2014){Arnalte-Mur}, {Mart{\'{\i}}nez},
  {Norberg}, {Fern{\'a}ndez-Soto}, {Ascaso}, {Merson}, {Aguerri}, {Castander},
  {Hurtado-Gil}, {L{\'o}pez-Sanjuan}, {Molino}, {Montero-Dorta}, {Stefanon},
  {Alfaro}, {Aparicio-Villegas}, {Ben{\'{\i}}tez}, {Broadhurst},
  {Cabrera-Ca{\~n}o}, {Cepa}, {Cervi{\~n}o}, {Crist{\'o}bal-Hornillos}, {del
  Olmo}, {Gonz{\'a}lez Delgado}, {Husillos}, {Infante}, {M{\'a}rquez},
  {Masegosa}, {Moles}, {Perea}, {Povi{\'c}}, {Prada}, \&
  {Quintana}}]{arnaltemur14}
{Arnalte-Mur}, P., {Mart{\'{\i}}nez}, V.~J., {Norberg}, P., {et~al.} 2014,
  \mnras, 441, 1783

\bibitem[{{Astropy Collaboration} {et~al.}(2013){Astropy Collaboration},
  {Robitaille}, {Tollerud}, {Greenfield}, {Droettboom}, {Bray}, {Aldcroft},
  {Davis}, {Ginsburg}, {Price-Whelan}, {Kerzendorf}, {Conley}, {Crighton},
  {Barbary}, {Muna}, {Ferguson}, {Grollier}, {Parikh}, {Nair}, {Unther},
  {Deil}, {Woillez}, {Conseil}, {Kramer}, {Turner}, {Singer}, {Fox}, {Weaver},
  {Zabalza}, {Edwards}, {Azalee Bostroem}, {Burke}, {Casey}, {Crawford},
  {Dencheva}, {Ely}, {Jenness}, {Labrie}, {Lim}, {Pierfederici}, {Pontzen},
  {Ptak}, {Refsdal}, {Servillat}, \& {Streicher}}]{astropy}
{Astropy Collaboration}, {Robitaille}, T.~P., {Tollerud}, E.~J., {et~al.} 2013,
  \aap, 558, A33

\bibitem[{{Azzalini}(2005)}]{azzalini05}
{Azzalini}, A. 2005, Scandinavian Journal of Statistics, 32, 159

\bibitem[{{Barro} {et~al.}(2011){Barro}, {P{\'e}rez-Gonz{\'a}lez}, {Gallego},
  {Ashby}, {Kajisawa}, {Miyazaki}, {Villar}, {Yamada}, \&
  {Zamorano}}]{barro11mass}
{Barro}, G., {P{\'e}rez-Gonz{\'a}lez}, P.~G., {Gallego}, J., {et~al.} 2011,
  \apjs, 193, 30

\bibitem[{{Bell} \& {de Jong}(2001)}]{bell01}
{Bell}, E.~F. \& {de Jong}, R.~S. 2001, \apj, 550, 212

\bibitem[{{Bell} {et~al.}(2003){Bell}, {McIntosh}, {Katz}, \&
  {Weinberg}}]{bell03}
{Bell}, E.~F., {McIntosh}, D.~H., {Katz}, N., \& {Weinberg}, M.~D. 2003, \apjs,
  149, 289

\bibitem[{{Ben{\'{\i}}tez}(2000)}]{benitez00}
{Ben{\'{\i}}tez}, N. 2000, \apj, 536, 571

\bibitem[{{Bongiorno} {et~al.}(2016){Bongiorno}, {Schulze}, {Merloni},
  {Zamorani}, {Ilbert}, {La Franca}, {Peng}, {Piconcelli}, {Mainieri},
  {Silverman}, {Brusa}, {Fiore}, {Salvato}, \& {Scoville}}]{bongiorno16}
{Bongiorno}, A., {Schulze}, A., {Merloni}, A., {et~al.} 2016, \aap, 588, A78

\bibitem[{{Bruzual} \& {Charlot}(2003)}]{bc03}
{Bruzual}, G. \& {Charlot}, S. 2003, \mnras, 344, 1000

\bibitem[{{Calzetti} {et~al.}(2000){Calzetti}, {Armus}, {Bohlin}, {Kinney},
  {Koornneef}, \& {Storchi-Bergmann}}]{calzetti00}
{Calzetti}, D., {Armus}, L., {Bohlin}, R.~C., {et~al.} 2000, \apj, 533, 682

\bibitem[{{Chabrier}(2003)}]{chabrier03}
{Chabrier}, G. 2003, \pasp, 115, 763

\bibitem[{{Chang} {et~al.}(2015){Chang}, {van der Wel}, {da Cunha}, \&
  {Rix}}]{chang15}
{Chang}, Y.-Y., {van der Wel}, A., {da Cunha}, E., \& {Rix}, H.-W. 2015, \apjs,
  219, 8

\bibitem[{{Conroy} \& {Gunn}(2010)}]{conroy10}
{Conroy}, C. \& {Gunn}, J.~E. 2010, \apj, 712, 833

\bibitem[{{Courteau} {et~al.}(2014){Courteau}, {Cappellari}, {de Jong},
  {Dutton}, {Emsellem}, {Hoekstra}, {Koopmans}, {Mamon}, {Maraston}, {Treu}, \&
  {Widrow}}]{courteau14}
{Courteau}, S., {Cappellari}, M., {de Jong}, R.~S., {et~al.} 2014, Reviews of
  Modern Physics, 86, 47

\bibitem[{{Crist{\'o}bal-Hornillos} {et~al.}(2009){Crist{\'o}bal-Hornillos},
  {Aguerri}, {Moles}, {Perea}, {Castander}, {Broadhurst}, {Alfaro},
  {Ben{\'{\i}}tez}, {Cabrera-Ca{\~n}o}, {Cepa}, {Cervi{\~n}o},
  {Fern{\'a}ndez-Soto}, {Delgado}, {Husillos}, {Infante}, {M{\'a}rquez},
  {Mart{\'{\i}}nez}, {Masegosa}, {del Olmo}, {Prada}, {Quintana}, \&
  {S{\'a}nchez}}]{cristobal09}
{Crist{\'o}bal-Hornillos}, D., {Aguerri}, J.~A.~L., {Moles}, M., {et~al.} 2009,
  \apj, 696, 1554

\bibitem[{{D{\'{\i}}az-Garc{\'{\i}}a}
  {et~al.}(2017){D{\'{\i}}az-Garc{\'{\i}}a}, {Cenarro}, {L{\'o}pez-Sanjuan},
  {Ferreras}, {Cervi{\~n}o}, {Fern{\'a}ndez-Soto}, {M{\'a}rquez}, {Povi{\'c}},
  {San Roman}, {Viironen}, {Moles}, {Crist{\'o}bal-Hornillos}, {Alfaro},
  {Aparicio-Villegas}, {Ben{\'{\i}}tez}, {Broadhurst}, {Cabrera-Ca{\~n}o},
  {Castander}, {Cepa}, {Gonz{\'a}lez Delgado}, {Husillos}, {Infante},
  {Aguerri}, {Masegosa}, {Molino}, {del Olmo}, {Perea}, {Prada}, {Quintana}, \&
  {Mart{\'{\i}}nez}}]{diazgarcia18uvj}
{D{\'{\i}}az-Garc{\'{\i}}a}, L.~A., {Cenarro}, A.~J., {L{\'o}pez-Sanjuan}, C.,
  {et~al.} 2017, \aap, submitted [arXiv:1711.10590]

\bibitem[{{D{\'{\i}}az-Garc{\'{\i}}a}
  {et~al.}(2018){D{\'{\i}}az-Garc{\'{\i}}a}, {Cenarro}, {L{\'o}pez-Sanjuan},
  {Ferreras}, {Fern{\'a}ndez-Soto}, {Gonz{\'a}lez Delgado}, {M{\'a}rquez},
  {Masegosa}, {San Roman}, {Viironen}, {Bonoli}, {Cervi{\~n}o}, {Moles},
  {Crist{\'o}bal-Hornillos}, {Alfaro}, {Aparicio-Villegas}, {Ben{\'{\i}}tez},
  {Broadhurst}, {Cabrera-Ca{\~n}o}, {Castander}, {Cepa}, {Husillos}, {Infante},
  {Aguerri}, {Mart{\'{\i}}nez}, {Molino}, {del Olmo}, {Perea}, {Prada}, \&
  {Quintana}}]{diazgarcia18sp}
{D{\'{\i}}az-Garc{\'{\i}}a}, L.~A., {Cenarro}, A.~J., {L{\'o}pez-Sanjuan}, C.,
  {et~al.} 2018, \aap, submitted [arXiv:1802.06813]

\bibitem[{{D{\'{\i}}az-Garc{\'{\i}}a}
  {et~al.}(2015){D{\'{\i}}az-Garc{\'{\i}}a}, {Cenarro}, {L{\'o}pez-Sanjuan},
  {Ferreras}, {Varela}, {Viironen}, {Crist{\'o}bal-Hornillos}, {Moles},
  {Mar{\'{\i}}n-Franch}, {Arnalte-Mur}, {Ascaso}, {Cervi{\~n}o}, {Gonz{\'a}lez
  Delgado}, {M{\'a}rquez}, {Masegosa}, {Molino}, {Povi{\'c}}, {Alfaro},
  {Aparicio-Villegas}, {Ben{\'{\i}}tez}, {Broadhurst}, {Cabrera-Ca{\~n}o},
  {Castander}, {Cepa}, {Fern{\'a}ndez-Soto}, {Husillos}, {Infante}, {Aguerri},
  {Mart{\'{\i}}nez}, {del Olmo}, {Perea}, {Prada}, {Quintana}, \&
  {Gruel}}]{diazgarcia15}
{D{\'{\i}}az-Garc{\'{\i}}a}, L.~A., {Cenarro}, A.~J., {L{\'o}pez-Sanjuan}, C.,
  {et~al.} 2015, \aap, 582, A14

\bibitem[{{Driver} {et~al.}(2011){Driver}, {Hill}, {Kelvin}, {Robotham},
  {Liske}, {Norberg}, {Baldry}, {Bamford}, {Hopkins}, {Loveday}, {Peacock},
  {Andrae}, {Bland-Hawthorn}, {Brough}, {Brown}, {Cameron}, {Ching}, {Colless},
  {Conselice}, {Croom}, {Cross}, {de Propris}, {Dye}, {Drinkwater}, {Ellis},
  {Graham}, {Grootes}, {Gunawardhana}, {Jones}, {van Kampen}, {Maraston},
  {Nichol}, {Parkinson}, {Phillipps}, {Pimbblet}, {Popescu}, {Prescott},
  {Roseboom}, {Sadler}, {Sansom}, {Sharp}, {Smith}, {Taylor}, {Thomas},
  {Tuffs}, {Wijesinghe}, {Dunne}, {Frenk}, {Jarvis}, {Madore}, {Meyer},
  {Seibert}, {Staveley-Smith}, {Sutherland}, \& {Warren}}]{gama}
{Driver}, S.~P., {Hill}, D.~T., {Kelvin}, L.~S., {et~al.} 2011, \mnras, 413,
  971

\bibitem[{{Fioc} \& {Rocca-Volmerange}(1997)}]{pegase}
{Fioc}, M. \& {Rocca-Volmerange}, B. 1997, \aap, 326, 950

\bibitem[{{Fitzpatrick}(1999)}]{fitzpatrick99}
{Fitzpatrick}, E.~L. 1999, \pasp, 111, 63

\bibitem[{{Foreman-Mackey} {et~al.}(2013){Foreman-Mackey}, {Hogg}, {Lang}, \&
  {Goodman}}]{emcee}
{Foreman-Mackey}, D., {Hogg}, D.~W., {Lang}, D., \& {Goodman}, J. 2013, \pasp,
  125, 306

\bibitem[{{Gallazzi} \& {Bell}(2009)}]{gallazzi09}
{Gallazzi}, A. \& {Bell}, E.~F. 2009, \apjs, 185, 253

\bibitem[{{Gallazzi} {et~al.}(2014){Gallazzi}, {Bell}, {Zibetti}, {Brinchmann},
  \& {Kelson}}]{gallazzi14}
{Gallazzi}, A., {Bell}, E.~F., {Zibetti}, S., {Brinchmann}, J., \& {Kelson},
  D.~D. 2014, \apj, 788, 72

\bibitem[{{Gallazzi} {et~al.}(2005){Gallazzi}, {Charlot}, {Brinchmann},
  {White}, \& {Tremonti}}]{gallazzi05}
{Gallazzi}, A., {Charlot}, S., {Brinchmann}, J., {White}, S.~D.~M., \&
  {Tremonti}, C.~A. 2005, \mnras, 362, 41

\bibitem[{{Goodman} \& {Weare}(2010)}]{goodman10}
{Goodman}, J. \& {Weare}, J. 2010, Comm. App. Math. Comp. Sci., 5, 65

\bibitem[{{Herrmann} {et~al.}(2016){Herrmann}, {Hunter}, {Zhang}, \&
  {Elmegreen}}]{herrmann16}
{Herrmann}, K.~A., {Hunter}, D.~A., {Zhang}, H.-X., \& {Elmegreen}, B.~G. 2016,
  \aj, 152, 177

\bibitem[{{Huertas-Company} {et~al.}(2016){Huertas-Company}, {Bernardi},
  {P{\'e}rez-Gonz{\'a}lez}, {Ashby}, {Barro}, {Conselice}, {Daddi}, {Dekel},
  {Dimauro}, {Faber}, {Grogin}, {Kartaltepe}, {Kocevski}, {Koekemoer}, {Koo},
  {Mei}, \& {Shankar}}]{huertas16}
{Huertas-Company}, M., {Bernardi}, M., {P{\'e}rez-Gonz{\'a}lez}, P.~G.,
  {et~al.} 2016, \mnras, 462, 4495

\bibitem[{Hunter(2007)}]{pylab}
Hunter, J.~D. 2007, Computing In Science \& Engineering, 9, 90

\bibitem[{{Ilbert} {et~al.}(2006){Ilbert}, {Lauger}, {Tresse}, {Buat},
  {Arnouts}, {Le F{\`e}vre}, {Burgarella}, {Zucca}, {Bardelli}, {Zamorani},
  {Bottini}, {Garilli}, {Le Brun}, {Maccagni}, {Picat}, {Scaramella},
  {Scodeggio}, {Vettolani}, {Zanichelli}, {Adami}, {Arnaboldi}, {Bolzonella},
  {Cappi}, {Charlot}, {Contini}, {Foucaud}, {Franzetti}, {Gavignaud}, {Guzzo},
  {Iovino}, {McCracken}, {Marano}, {Marinoni}, {Mathez}, {Mazure}, {Meneux},
  {Merighi}, {Paltani}, {Pello}, {Pollo}, {Pozzetti}, {Radovich}, {Bondi},
  {Bongiorno}, {Busarello}, {Ciliegi}, {Mellier}, {Merluzzi}, {Ripepi}, \&
  {Rizzo}}]{ilbert06}
{Ilbert}, O., {Lauger}, S., {Tresse}, L., {et~al.} 2006, \aap, 453, 809

\bibitem[{{Into} \& {Portinari}(2013)}]{into13}
{Into}, T. \& {Portinari}, L. 2013, \mnras, 430, 2715

\bibitem[{{Jablonka} \& {Arimoto}(1992)}]{jablonka92}
{Jablonka}, J. \& {Arimoto}, N. 1992, \aap, 255, 63

\bibitem[{{Kauffmann} {et~al.}(2003){Kauffmann}, {Heckman}, {White}, {Charlot},
  {Tremonti}, {Peng}, {Seibert}, {Brinkmann}, {Nichol}, {SubbaRao}, \&
  {York}}]{kauffmann03}
{Kauffmann}, G., {Heckman}, T.~M., {White}, S.~D.~M., {et~al.} 2003, \mnras,
  341, 54

\bibitem[{{Lara-L{\'o}pez} {et~al.}(2010){Lara-L{\'o}pez}, {Cepa},
  {Bongiovanni}, {P{\'e}rez Garc{\'{\i}}a}, {Ederoclite}, {Casta{\~n}eda},
  {Fern{\'a}ndez Lorenzo}, {Povi{\'c}}, \& {S{\'a}nchez-Portal}}]{lara10}
{Lara-L{\'o}pez}, M.~A., {Cepa}, J., {Bongiovanni}, A., {et~al.} 2010, \aap,
  521, L53

\bibitem[{{L{\'o}pez-Sanjuan} {et~al.}(2017){L{\'o}pez-Sanjuan}, {Tempel},
  {Ben{\'{\i}}tez}, {Molino}, {Viironen}, {D{\'{\i}}az-Garc{\'{\i}}a},
  {Fern{\'a}ndez-Soto}, {Santos}, {Varela}, {Cenarro}, {Moles}, {Arnalte-Mur},
  {Ascaso}, {Montero-Dorta}, {Povi{\'c}}, {Mart{\'{\i}}nez}, {Nieves-Seoane},
  {Stefanon}, {Hurtado-Gil}, {M{\'a}rquez}, {Perea}, {Aguerri}, {Alfaro},
  {Aparicio-Villegas}, {Broadhurst}, {Cabrera-Ca{\~n}o}, {Castander}, {Cepa},
  {Cervi{\~n}o}, {Crist{\'o}bal-Hornillos}, {Gonz{\'a}lez Delgado}, {Husillos},
  {Infante}, {Masegosa}, {del Olmo}, {Prada}, \& {Quintana}}]{clsj17lfbal}
{L{\'o}pez-Sanjuan}, C., {Tempel}, E., {Ben{\'{\i}}tez}, N., {et~al.} 2017,
  \aap, 599, A62

\bibitem[{{Mannucci} {et~al.}(2009){Mannucci}, {Cresci}, {Maiolino}, {Marconi},
  {Pastorini}, {Pozzetti}, {Gnerucci}, {Risaliti}, {Schneider}, {Lehnert}, \&
  {Salvati}}]{mannucci09}
{Mannucci}, F., {Cresci}, G., {Maiolino}, R., {et~al.} 2009, \mnras, 398, 1915

\bibitem[{{Maraston}(2005)}]{maraston05}
{Maraston}, C. 2005, \mnras, 362, 799

\bibitem[{{McGaugh} \& {Schombert}(2014)}]{mcgaugh14}
{McGaugh}, S.~S. \& {Schombert}, J.~M. 2014, \aj, 148, 77

\bibitem[{{Moffett} {et~al.}(2016){Moffett}, {Ingarfield}, {Driver},
  {Robotham}, {Kelvin}, {Lange}, {Me{\v s}tri{\'c}}, {Alpaslan}, {Baldry},
  {Bland-Hawthorn}, {Brough}, {Cluver}, {Davies}, {Holwerda}, {Hopkins},
  {Kafle}, {Kennedy}, {Norberg}, \& {Taylor}}]{moffett16}
{Moffett}, A.~J., {Ingarfield}, S.~A., {Driver}, S.~P., {et~al.} 2016, \mnras,
  457, 1308

\bibitem[{{Moles} {et~al.}(2008){Moles}, {Ben{\'{\i}}tez}, {Aguerri}, {Alfaro},
  {Broadhurst}, {Cabrera-Ca{\~n}o}, {Castander}, {Cepa}, {Cervi{\~n}o},
  {Crist{\'o}bal-Hornillos}, {Fern{\'a}ndez-Soto}, {Gonz{\'a}lez Delgado},
  {Infante}, {M{\'a}rquez}, {Mart{\'{\i}}nez}, {Masegosa}, {del Olmo}, {Perea},
  {Prada}, {Quintana}, \& {S{\'a}nchez}}]{alhambra}
{Moles}, M., {Ben{\'{\i}}tez}, N., {Aguerri}, J.~A.~L., {et~al.} 2008, \aj,
  136, 1325

\bibitem[{{Molino} {et~al.}(2014){Molino}, {Ben{\'{\i}}tez}, {Moles},
  {Fern{\'a}ndez-Soto}, {Crist{\'o}bal-Hornillos}, {Ascaso},
  {Jim{\'e}nez-Teja}, {Schoenell}, {Arnalte-Mur}, {Povi{\'c}}, {Coe},
  {L{\'o}pez-Sanjuan}, {D{\'{\i}}az-Garc{\'{\i}}a}, {Varela}, {Stefanon},
  {Cenarro}, {Matute}, {Masegosa}, {M{\'a}rquez}, {Perea}, {Del Olmo},
  {Husillos}, {Alfaro}, {Aparicio-Villegas}, {Cervi{\~n}o}, {Huertas-Company},
  {Aguerri}, {Broadhurst}, {Cabrera-Ca{\~n}o}, {Cepa}, {Gonz{\'a}lez},
  {Infante}, {Mart{\'{\i}}nez}, {Prada}, \& {Quintana}}]{molino13}
{Molino}, A., {Ben{\'{\i}}tez}, N., {Moles}, M., {et~al.} 2014, \mnras, 441,
  2891

\bibitem[{{Montero-Dorta} {et~al.}(2016){Montero-Dorta}, {Bolton},
  {Brownstein}, {Swanson}, {Dawson}, {Prada}, {Eisenstein}, {Maraston},
  {Thomas}, {Comparat}, {Chuang}, {McBride}, {Favole}, {Guo},
  {Rodr{\'{\i}}guez-Torres}, \& {Schneider}}]{monterodorta16lf}
{Montero-Dorta}, A.~D., {Bolton}, A.~S., {Brownstein}, J.~R., {et~al.} 2016,
  \mnras, 461, 1131

\bibitem[{{Nieves-Seoane} {et~al.}(2017){Nieves-Seoane}, {Fernandez-Soto},
  {Arnalte-Mur}, {Molino}, {Stefanon}, {Ferreras}, {Ascaso}, {Ballesteros},
  {Crist{\'o}bal-Hornillos}, {L{\'o}pez-Sanju{\'a}n}, {Hurtado-Gil},
  {M{\'a}rquez}, {Masegosa}, {Aguerri}, {Alfaro}, {Aparicio-Villegas},
  {Ben{\'{\i}}tez}, {Broadhurst}, {Cabrera-Ca{\~n}o}, {Castander}, {Cepa},
  {Cervi{\~n}o}, {Gonz{\'a}lez Delgado}, {Husillos}, {Infante},
  {Mart{\'{\i}}nez}, {Moles}, {Olmo}, {Perea}, {Povi{\'c}}, {Prada},
  {Quintana}, {Troncoso-Iribarren}, \& {Viironen}}]{nieves17}
{Nieves-Seoane}, L., {Fernandez-Soto}, A., {Arnalte-Mur}, P., {et~al.} 2017,
  \mnras, 464, 4331

\bibitem[{{Noeske} {et~al.}(2007){Noeske}, {Weiner}, {Faber}, {Papovich},
  {Koo}, {Somerville}, {Bundy}, {Conselice}, {Newman}, {Schiminovich}, {Le
  Floc'h}, {Coil}, {Rieke}, {Lotz}, {Primack}, {Barmby}, {Cooper}, {Davis},
  {Ellis}, {Fazio}, {Guhathakurta}, {Huang}, {Kassin}, {Martin}, {Phillips},
  {Rich}, {Small}, {Willmer}, \& {Wilson}}]{noeske07}
{Noeske}, K.~G., {Weiner}, B.~J., {Faber}, S.~M., {et~al.} 2007, \apjl, 660,
  L43

\bibitem[{{Oke} \& {Gunn}(1983)}]{oke83}
{Oke}, J.~B. \& {Gunn}, J.~E. 1983, \apj, 266, 713

\bibitem[{{Portinari} {et~al.}(2004){Portinari}, {Sommer-Larsen}, \&
  {Tantalo}}]{portinari04}
{Portinari}, L., {Sommer-Larsen}, J., \& {Tantalo}, R. 2004, \mnras, 347, 691

\bibitem[{{Roediger} \& {Courteau}(2015)}]{roediger15}
{Roediger}, J.~C. \& {Courteau}, S. 2015, \mnras, 452, 3209

\bibitem[{Schwarz(1978)}]{schwarz78}
Schwarz, G. 1978, Ann. Statist., 6, 461

\bibitem[{{Shen} {et~al.}(2003){Shen}, {Mo}, {White}, {Blanton}, {Kauffmann},
  {Voges}, {Brinkmann}, \& {Csabai}}]{shen03}
{Shen}, S., {Mo}, H.~J., {White}, S.~D.~M., {et~al.} 2003, \mnras, 343, 978

\bibitem[{{Taylor} {et~al.}(2015){Taylor}, {Hopkins}, {Baldry},
  {Bland-Hawthorn}, {Brown}, {Colless}, {Driver}, {Norberg}, {Robotham},
  {Alpaslan}, {Brough}, {Cluver}, {Gunawardhana}, {Kelvin}, {Liske},
  {Conselice}, {Croom}, {Foster}, {Jarrett}, {Lara-Lopez}, \&
  {Loveday}}]{taylor15}
{Taylor}, E.~N., {Hopkins}, A.~M., {Baldry}, I.~K., {et~al.} 2015, \mnras, 446,
  2144

\bibitem[{{Taylor} {et~al.}(2011){Taylor}, {Hopkins}, {Baldry}, {Brown},
  {Driver}, {Kelvin}, {Hill}, {Robotham}, {Bland-Hawthorn}, {Jones}, {Sharp},
  {Thomas}, {Liske}, {Loveday}, {Norberg}, {Peacock}, {Bamford}, {Brough},
  {Colless}, {Cameron}, {Conselice}, {Croom}, {Frenk}, {Gunawardhana},
  {Kuijken}, {Nichol}, {Parkinson}, {Phillipps}, {Pimbblet}, {Popescu},
  {Prescott}, {Sutherland}, {Tuffs}, {van Kampen}, \& {Wijesinghe}}]{taylor11}
{Taylor}, E.~N., {Hopkins}, A.~M., {Baldry}, I.~K., {et~al.} 2011, \mnras, 418,
  1587

\bibitem[{{Tinsley}(1981)}]{tinsley81}
{Tinsley}, B.~M. 1981, \mnras, 194, 63

\bibitem[{{Tremonti} {et~al.}(2004){Tremonti}, {Heckman}, {Kauffmann},
  {Brinchmann}, {Charlot}, {White}, {Seibert}, {Peng}, {Schlegel}, {Uomoto},
  {Fukugita}, \& {Brinkmann}}]{tremonti04}
{Tremonti}, C.~A., {Heckman}, T.~M., {Kauffmann}, G., {et~al.} 2004, \apj, 613,
  898

\bibitem[{{Trujillo} {et~al.}(2007){Trujillo}, {Conselice}, {Bundy}, {Cooper},
  {Eisenhardt}, \& {Ellis}}]{trujillo07}
{Trujillo}, I., {Conselice}, C.~J., {Bundy}, K., {et~al.} 2007, \mnras, 382,
  109

\bibitem[{{van de Sande} {et~al.}(2015){van de Sande}, {Kriek}, {Franx},
  {Bezanson}, \& {van Dokkum}}]{vandesande15}
{van de Sande}, J., {Kriek}, M., {Franx}, M., {Bezanson}, R., \& {van Dokkum},
  P.~G. 2015, \apj, 799, 125

\bibitem[{{van der Wel} {et~al.}(2014){van der Wel}, {Franx}, {van Dokkum},
  {Skelton}, {Momcheva}, {Whitaker}, {Brammer}, {Bell}, {Rix}, {Wuyts},
  {Ferguson}, {Holden}, {Barro}, {Koekemoer}, {Chang}, {McGrath},
  {H{\"a}ussler}, {Dekel}, {Behroozi}, {Fumagalli}, {Leja}, {Lundgren},
  {Maseda}, {Nelson}, {Wake}, {Patel}, {Labb{\'e}}, {Faber}, {Grogin}, \&
  {Kocevski}}]{vanderwel14}
{van der Wel}, A., {Franx}, M., {van Dokkum}, P.~G., {et~al.} 2014, \apj, 788,
  28

\bibitem[{{Vazdekis} {et~al.}(2016){Vazdekis}, {Koleva}, {Ricciardelli},
  {R{\"o}ck}, \& {Falc{\'o}n-Barroso}}]{vazdekis16}
{Vazdekis}, A., {Koleva}, M., {Ricciardelli}, E., {R{\"o}ck}, B., \&
  {Falc{\'o}n-Barroso}, J. 2016, \mnras, 463, 3409

\bibitem[{{Viironen} {et~al.}(2017){Viironen}, {L{\'o}pez-Sanjuan},
  {Hern{\'a}ndez-Monteagudo}, {Chaves-Montero}, {Ascaso}, {Bonoli},
  {Crist{\'o}bal-Hornillos}, {D{\'{\i}}az-Garc{\'{\i}}a}, {Fern{\'a}ndez-Soto},
  {M{\'a}rquez}, {Masegosa}, {Povi{\'c}}, {Varela}, {Cenarro}, {Aguerri},
  {Alfaro}, {Aparicio-Villegas}, {Ben{\'{\i}}tez}, {Broadhurst},
  {Cabrera-Ca{\~n}o}, {Castander}, {Cepa}, {Cervi{\~n}o}, {Gonz{\'a}lez
  Delgado}, {Husillos}, {Infante}, {Mart{\'{\i}}nez}, {Moles}, {Molino}, {del
  Olmo}, {Perea}, {Prada}, \& {Quintana}}]{viironen18}
{Viironen}, K., {L{\'o}pez-Sanjuan}, C., {Hern{\'a}ndez-Monteagudo}, C.,
  {et~al.} 2017, \aap, in press [arXiv:1712.01028]

\bibitem[{{Williams} {et~al.}(2009){Williams}, {Quadri}, {Franx}, {van Dokkum},
  \& {Labb{\'e}}}]{williams09}
{Williams}, R.~J., {Quadri}, R.~F., {Franx}, M., {van Dokkum}, P., \&
  {Labb{\'e}}, I. 2009, \apj, 691, 1879

\bibitem[{{Zaritsky} {et~al.}(2014){Zaritsky}, {Gil de Paz}, \&
  {Bouquin}}]{zaritsky14}
{Zaritsky}, D., {Gil de Paz}, A., \& {Bouquin}, A.~Y.~K. 2014, \apjl, 780, L1

\bibitem[{{Zibetti} {et~al.}(2009){Zibetti}, {Charlot}, \& {Rix}}]{zibetti09}
{Zibetti}, S., {Charlot}, S., \& {Rix}, H.-W. 2009, \mnras, 400, 1181

\end{thebibliography}
\bibliographystyle{aa}

\end{document}